\documentclass[12pt]{iopart}
\usepackage{iopams}  
\usepackage{graphicx}

\begin{document}

\title[Weak measurement of photon polarization]
{Weak measurement of photon polarization 
by back-action induced path interference}

\author{Masataka~Iinuma$^1$, Yutaro~Suzuki$^1$, Gen~Taguchi$^1$, Yutaka~Kadoya$^1$, 
and Holger~F.~Hofmann$^{1,2}$}

\address{$^1$ Graduate school of Advanced Sciences of Matter, Hiroshima University \\
 1-3-1 Kagamiyama, Higashi-Hiroshima, Japan, 739-8530, \\
$^2$ JST,Crest, Sanbancho 5, Chiyoda-ku, Tokyo 102-0075, Japan
}
\ead{iinuma@hiroshima-u.ac.jp}

\begin{abstract}
The essential feature of weak measurements on quantum systems is the reduction of measurement back-action to negligible levels. To observe the non-classical features of weak measurements, it is therefore more important to avoid additional back-action errors than it is to avoid errors in the actual measurement outcome. In this paper, it is shown how an optical weak measurement of diagonal (PM) polarization can be realized by path interference between the horizontal (H) and vertical (V) polarization components of the input beam. The measurement strength can then be controlled by rotating the H and V polarizations towards each other. This well-controlled operation effectively generates the back-action without additional decoherence, while the visibility of the interference between the two beams only limits the measurement resolution. As the experimental results confirm, we can obtain extremely high weak values, even at rather low visibilities. Our method therefore provides a realization of weak measurements that is extremely robust against experimental imperfections.
\end{abstract}


\vspace{1cm}

submitted to {\it New J. Phys}

\maketitle

\section{Introduction}

In ideal quantum measurements, there is a trade-off between the information 
obtained about the measured observable and the back-action suffered by observables that do not share any eigenstates with the measured observable. A fully resolved strong measurement has a maximal back-action since it completely removes any coherences between the eigenstates of the measured observable. On the other hand, a weak measurement with low resolution can have negligible back-action, leaving the coherences of the initial state almost completely intact. As first pointed out by Aharonov et al. \cite{Aharonov88}, it is then possible to obtain measurement results far outside the spectrum of the eigenvalues of the measured observable by post-selecting a specific final measurement outcome. In the limit of negligible back-action, these post-selected results only depend on the initial state, the final state, and the operator of the measured observable. It is therefore possible to define the measurement result as the weak value of the measured observable for the specific combination of initial state and final state defined by state preparation and post-selection.  

It was soon realized that photon polarization was an ideal system for the experimental realization of weak measurements, since optics provides optimal control of coherence using well-established technologies \cite{Duck89,Ritchie91}. At first, the implications and the usefulness of weak values were unclear. However, recent advances in quantum technologies have revived the interest in the unusual properties of weak values, with possible applications in precision measurements \cite{Hosten08,Dixon09,Bru10,Hof10a}, realizations using quantum logic gates \cite{Pryde05,Ralph06}, resolution of quantum paradoxes \cite{Resch04,Wil08,Lundeen09,Yokota09,Goggin09} and more fundamental implications for quantum statistics and quantum physics \cite{Tamate09,Hofmann10,Hosoya10}. Because of the wide range of problems that can be addressed by weak measurements, it seems to be desirable to develop simple and efficient technological implementations that are not too sensitive to experimental errors. In the following, we therefore present an experimental setup for the weak measurement of photon polarization that uses a basic two path interferometer as meter system. In such a setup, the essential problem is the limited visibility of the interferences between the two paths. If the two paths correspond to the eigenstates of the measured observable, the limited visibility causes an additional back-action on coherent superpositions of these eigenstates, limiting the magnitude of the weak values observed in the experiment. We therefore propose an alternative realization of weak measurements where the interference occurs between eigenstates of the back-action observable defined by the post-selection. As we discuss below, it is then possible to control the back-action precisely by implementing it separately in each path. The measurement effect is then obtained because this back-action induces an interference effect that depends on the quantum coherences between the back-action observable. The result is a conventional weak measurement (or variable strength measurement), but now the errors caused by finite visibility of the interference only reduce the measurement resolution, without causing any additional back-action. Our setup is therefore ideally suited for weak measurements in the presence of experimental imperfections.

The rest of the paper is organized as follows. In sec. \ref{sec:setup} we describe the principle of back-action induced interference and the experimental setup used to realize it. In sec. \ref{sec:results}, we present the experimental results obtained for the weak values of photon polarization and show that the errors are close to the theoretical limit for the measurement strength used in the experiment. In sec. \ref{sec:urel}, we present experimental results for the trade-off between resolution and back-action in our setup. The results show that the visibility only limits measurement resolution, without contributing to the back-action. Sec. \ref{sec:conclusion} summarizes the results and concludes the paper.

\section{Quantum measurement by back-action induced interference} 
\label{sec:setup}

In our experiment, we realize a measurement of photon polarization with variable measurement strength by making use of the fact that the diagonal polarization is determined by the phase coherence between the horizontal (H) and vertical (V) polarizations. The positive (P) and the negative (M) superpositions can therefore be distinguished by interference between the H and V components of the photon state. Although path interference between these two components cannot occur if the beams corresponding to the H and V polarizations can still be distinguished by their orthogonal polarization states, it is possible to induce a well-controlled amount of interference by erasing the HV information in the beams using a coherent rotation of polarization towards a common diagonal polarization. The increase of interference as the polarizations become less and less distinguishable corresponds to the trade-off between measurement information and back-action in the quantum measurement. Significantly, the final interference that results in a correlation between the output path and the PM polarization of the input state does not change the HV polarization at all, regardless of the visibility of the interference. The flips of HV polarization caused by the measurement back-action are therefore limited to the flips caused by the rotation of the polarization in the two arms of the interferometer. This method thus ensures optimal control of the back-action, permitting arbitrarily low back-action even in the presence of significant experimental errors. 

\begin{figure}[h]
\centerline{\includegraphics[width=100mm]{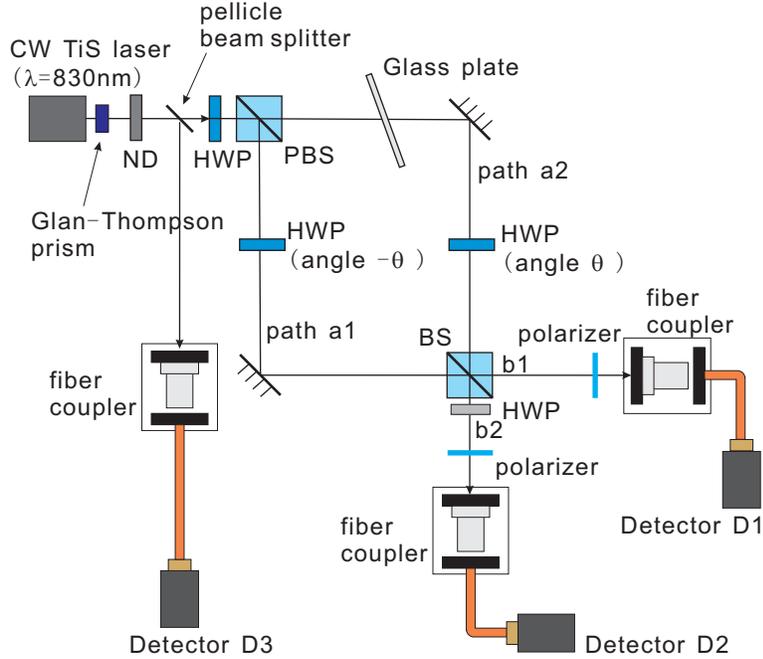}}
\caption{Setup of the polarization measurement using back-action induced interference. Interferences between the V polarized path a1 and the H polarized path a2 is induced using oppositely rotated HWP to reduce the angle between the polarization and therefore the distinguishability of the paths in terms of polarization. The output beams then distinguish positive and negative superpositions of H and V, corresponding to diagonal P and M polarization.}
\label{fig:setup}
\end{figure}

Figure \ref{fig:setup} shows our setup in more detail. 
The photon path is split by a polarizing beam splitter (PBS) 
into a V polarized path a1 and an H polarized path a2. The polarizations are then rotated in opposite directions by half-wave-plates (HWP) mounted in each path. Specifically, the HWP in path a1 is rotated by an angle of $- \theta $ from the horizontal/vertical alignment, whereas the HWP in path a2 is rotated by an angle of $ + \theta $. Finally, the two polarization components interfere at a beam splitter (BS) with 50 \% reflectivity for all polarizations, resulting in the output beams b1 and b2. A glass plate is used to compensate path length differences between path a1 and path a2, and a HWP is inserted in b2 to compensate for the phase shift caused by the difference in the number of reflections between the H and V components. 

The input photons were generated by using a CW titanium-sapphire laser(wavelength 830 nm, output power 300 mW) and passed through a Glan-Thompson prism to select photons with horizontal polarization. Neutral density(ND) filters were used to obtain intensities suitable for single photon counting with typical count rates around 100 kHz. The initial state of photon polarization was prepared by rotating the HWP upstream of the PBS. The number of output photons in path b1 and path b2 were counted using the single photon counting modules ( SPCM-AQR-14 ) D1 and D2, which were optically coupled to the paths b1 and b2 through fiber couplers and optical fibers. To keep track of fluctuations in the input light, the input intensity was monitored by detecting photons reflected by a pellicle beam splitter( reflectivity=8\% ) with another single photon counting module D3. Experimentally, the ratio of counts in D3 to total counts in D1 and D2 was found to be 0.020. Post-selection was realized by inserting polarizers in the output beams to select only the desired output polarization in both paths. 

The weak measurement is realized by the interferometer setup between the PBS and the BS.
If the polarizations in path a1 and a2 are orthogonal at the final beam splitter, no interference will be observed in the output probabilities of $b1$ and $b2$. The interference at the final beam splitter will simply restore the original superposition of H and V polarization of the input state. By rotating the polarization in both arms towards the same diagonal polarization P, the distinguishability of the two paths is reduced and the phase coherence between the HV components is converted into interferences between the paths. As a result, the probability of finding the photon in path b1 increases for positive superpositions of H and V (P polarization), and decreases for negative superpositions (M polarization). 
In the absence of experimental imperfections, the interference at the final beam splitter can be expressed in terms of the polarization vectors in paths a1 and a2 given in the $HV$-basis,
\begin{eqnarray}
\mid b1 \rangle & = & \frac{1}{\sqrt{2}}
C_{\mathrm{H}}
\left [
\begin{array}{l}
\cos 2 \theta \\
\sin 2 \theta 
\end{array}
\right ]
+
\frac{1}{\sqrt{2}}
C_{\mathrm{V}}
\left [
\begin{array}{l}
\sin 2 \theta \\
\cos 2 \theta 
\end{array}
\right ] \nonumber 
\\
\mid b2^\prime \rangle & = & \frac{1}{\sqrt{2}}
C_{\mathrm{H}}
\left [
\begin{array}{l}
\cos 2 \theta \\
\sin 2 \theta 
\end{array}
\right ]
-
\frac{1}{\sqrt{2}}
C_{\mathrm{V}}
\left [
\begin{array}{l}
\sin 2 \theta \\
\cos 2 \theta 
\end{array}
\right ] 
,
\end{eqnarray}
where $C_{\mathrm{H}}$ and $C_{\mathrm{V}}$ are the probability amplitudes of the H and V polarized components of the input state. For $\theta=0$, $\mid b1 \rangle$ reproduces the input polarization, while the polarization in $\mid b2^\prime \rangle$ is changed by a phase flip between the H and the V components. The HWP in b2 compensates this phase flip, resulting in the non-normalized output states $\mid b1 \rangle$ and $\mid b2 \rangle$ in the output beams of the measurement setup, 
\begin{eqnarray}
\mid b1 \rangle = 
\frac{1}{\sqrt{2}}
\left [
\begin{array}{rr}
\cos 2 \theta & \: \: \: \: \sin 2 \theta \\
\: \: \: \: \sin 2 \theta & \cos 2 \theta
\end{array}
\right ]
\left [ 
\begin{array}{c}
C_{\mathrm{H}} \\
C_{\mathrm{V}}
\end{array}
\right ] 
& = & 
\hat{M}_{\mathrm{b1}} \, | \psi_{i} \rangle
\nonumber
\\
\mid b2 \rangle =  
\frac{1}{\sqrt{2}}
\left [
\begin{array}{rr}
\cos 2 \theta & -\sin 2 \theta \\
-\sin 2 \theta & \cos 2 \theta
\end{array}
\right ]
\left [ 
\begin{array}{c}
C_{\mathrm{H}} \\
C_{\mathrm{V}}
\end{array}
\right ]
&=& \hat{M}_{\mathrm{b2}} \, | \psi_i \rangle, 
\label{eqn:b12}
\end{eqnarray}
where $ \mid \psi_i \rangle $ is the input state defined by $C_{\mathrm{H}} $ and $C_{\mathrm{V}}$ and the measurement operators $\hat{M}_{\mathrm{b1}}$ and $\hat{M}_{\mathrm{b2}}$ represent the effects of the measurement described by their matrix representation in the $HV$-basis. 

It is easy to see that the eigenstates of the measurement operators are the positive and negative superpositions of $\mid H \rangle$ and $\mid V \rangle$, 
corresponding to the diagonal polarization states, $| P \rangle$ and $| M \rangle$. In terms of the Stokes parameter $\hat{S}_{\mathrm{PM}} = | P \rangle \langle P | - | M \rangle \langle M | $, the positive operator measure defining the probabilities of finding the photon in b1 or b2 is therefore given by
\begin{eqnarray}
\hat{M}_{\mathrm{b1}}^{\dagger} \hat{M}_{\mathrm{b1}} &=& \frac{1}{2}
\left (
\hat{1} + \epsilon_{\mathrm{PM}} \, \hat{S}_{\mathrm{PM}}
\right ) \nonumber
\\
\hat{M}_{\mathrm{b2}}^{\dagger} \hat{M}_{\mathrm{b2}} &=& \frac{1}{2}
\left (
\hat{1} - \epsilon_{\mathrm{PM}} \, \hat{S}_{\mathrm{PM}}
\right ), 
\label{eqn:eb12}
\end{eqnarray}
where $\epsilon_{\mathrm{PM}} = \sin 4 \theta $ determines the measurement resolution. 
Without post-selection, the difference between the output probabilities in b1 and b2 is directly related to the PM polarization of the input state,
\begin{eqnarray} 
P(\mathrm{b1}) - P(\mathrm{b2}) 
& = & 
\langle \psi_{i} | 
{\hat{M}_{\mathrm{b1}}}^{\dagger} \hat{M}_{\mathrm{b1}} 
| \psi_{i} \rangle 
-
\langle \psi_{i} |
{\hat{M}_{\mathrm{b2}}}^{\dagger} \hat{M}_{\mathrm{b2}} 
| \psi_{i} \rangle \nonumber \\
& = & 
\epsilon_{\mathrm{PM}} \, \langle \psi_{i} | \hat{S}_{\mathrm{PM}} | \psi_{i} \rangle. 
\label{eqn:nops}
\end{eqnarray}
Experimentally, it is therefore possible to determine the polarization of the input light by dividing the difference in output probability by a constant value $\epsilon_{\mathrm{PM}}$, where the proper value of $\epsilon_{\mathrm{PM}}$ can be determined from the visibility obtained for maximally polarized inputs. 

In the case of output post-selection, the difference of the conditional output probabilities can now be interpreted as a conditional measurement of PM polarization. Experimentally, the conditional value $ \langle \hat{S}_{\mathrm{PM}} \rangle_{\mathrm{exp}} (m_f)$ obtained by post-selecting an output polarization state $\mid m_f \rangle$ is determined from the output probabilities by 
\begin{equation}
\langle \hat{S}_{\mathrm{PM}} \rangle_{\mathrm{exp}} (m_f) = \frac{1}{\epsilon_{\mathrm{PM}}}
\left ( P(\mathrm{b1} | \mathrm{m}_{f}) - P(\mathrm{b2} | \mathrm{m}_{f}) \right ).
\label{eqn:def-con}
\end{equation}
In the limit of negligible back-action ($\epsilon_{\mathrm{PM}} \to 0$), this experimental value is equal to the theoretically predicted weak value,
\begin{equation}
\langle \hat{S}_{\mathrm{PM}} \rangle_{\mathrm{weak}} = \mathrm{Re}
\left [
\frac{\langle \mathrm{m}_{f} | \hat{S}_{\mathrm{PM}} | \psi_{i} \rangle}
{\langle \mathrm{m}_{f} | \psi_{i} \rangle}
\right ].
\end{equation}
However, the finite measurement back-action for non-zero measurement resolutions $\epsilon_{\mathrm{PM}}$ modifies this result even in the case of an ideal measurement. Using Eq. (\ref{eqn:def-con}) to determine the conditional probabilities, the experimental value expected at finite back-action is
\begin{eqnarray}
\langle \hat{S}_{\mathrm{PM}} \rangle_{\mathrm{exp}} (m_f) 
& = & \frac{1}{\epsilon_{\mathrm{PM}}} 
\frac{| \langle \mathrm{m}_{f} | \hat{M}_{\mathrm{b1}} | \psi_{i} \rangle |^{2} - 
| \langle \mathrm{m}_{f} | \hat{M}_{\mathrm{b2}} | \psi_{i} \rangle |^{2}}
{| \langle \mathrm{m}_{f} | \hat{M}_{\mathrm{b1}} | \psi_{i} \rangle |^{2} + 
| \langle \mathrm{m}_{f} | \hat{M}_{\mathrm{b2}} | \psi_{i} \rangle |^{2}} \nonumber \\
 & = & 
\frac{| \langle \mathrm{m}_{f} | \psi_{i} \rangle |^{2}}{| \langle \mathrm{m}_{f} | \psi_{i} \rangle |^{2}+\eta_{\mathrm{HV}} \Delta_{\mathrm{flip}}}
\langle \hat{S}_{\mathrm{PM}} \rangle_{\mathrm{weak}}
\label{eqn:ps}
\end{eqnarray}
where $\eta_{\mathrm{HV}}=\sin^2(2\theta)$ is equal to the transition probability between H and V polarization given by the measurement operators $\hat{M}_{\mathrm{b1}}$ and $\hat{M}_{\mathrm{b2}}$, and $\Delta_{\mathrm{flip}}$ is the change in the post-selection probability caused by a polarization flip described by the operator $\hat{S}_{\mathrm{PM}}$, given by
\begin{equation}
\Delta_{\mathrm{flip}} = 
| \langle \mathrm{m}_{f} | \hat{S}_{\mathrm{PM}} | \psi_{i} \rangle |^{2} - | \langle \mathrm{m}_{f} | \psi_{i} \rangle |^{2}.
\end{equation}

As shown in Eq. (\ref{eqn:ps}), the experimental value is approximately equal to the weak value if the back-action induced change in the post-selection probability given by $\eta_{\mathrm{HV}} \Delta_{\mathrm{flip}}$ is sufficiently smaller than the original post-selection probability of $| \langle \mathrm{m}_{f} | \psi_{i} \rangle |^{2}$. However, extremely large weak values can only be obtained when the original post-selection probability goes to zero. To achieve extremely enhanced experimental weak values, it is therefore essential to keep the transition probability $\eta_{\mathrm{HV}}$ as small as possible. In particular, it is necessary to avoid additional errors from dephasing between the P and M polarized components. In our setup, we achieve extremely small values of $\eta_{\mathrm{HV}}$ by limiting the use of path interferences to an interference between a path associated with the initial H polarization and a path associated with the initial V polarization, therefore avoiding the HV transitions that would be caused by finite visibility interferences between the P and M polarized eigenstates of the measurement operators. As a result, our setup enables us to measure extremely high weak values, even at low visibilities of the path interference.

\section{Experimental demonstration of the weak measurement}
\label{sec:results}

For the experimental demonstration of the weak measurement, we chose a variable input state given by $C_{\mathrm{H}}=\sin \phi$ and $C_{\mathrm{V}}=\cos \phi$. Post-selection was implemented by inserting polarization filters selecting only the H polarized output components between the output ports and the detectors. Ideally, we should then be able to observe a theoretical weak value of $\langle \hat{S}_{\mathrm{PM}}\rangle_{\mathrm{weak}}=1/\tan \phi$. However, the measurement back-action modifies the directly determined experimental values to
\begin{equation}
\langle \hat{S}_{\mathrm{PM}} \rangle_{\mathrm{exp}} = 
\frac{\sin\phi \cos\phi}{\sin^2 \phi + \eta_{\mathrm{HV}} (\cos^2\phi-\sin^2 \phi)},
\label{eqn:exp-wv}
\end{equation}
where $\eta_{\mathrm{HV}}$ is the transition probability between H and V polarizations, including both the uncertainty limited back-action and additional effects of experimental imperfections in the setup. As discussed in the previous section, the back-action effects summarized by  $\eta_{\mathrm{HV}}$ limit the magnitude of the experimental weak values that can be observed experimentally. 
For small $\eta_{\mathrm{HV}}$, the maximal value is $\langle \hat{S}_{\mathrm{PM}} \rangle_{\mathrm{exp}}=1/\sqrt{4 \eta_{\mathrm{HV}}}$, obtained at an input polarization angle of $\phi=\sqrt{\eta_{\mathrm{HV}}}$. 

\begin{figure}[h]
\centerline{\includegraphics[width=100mm]{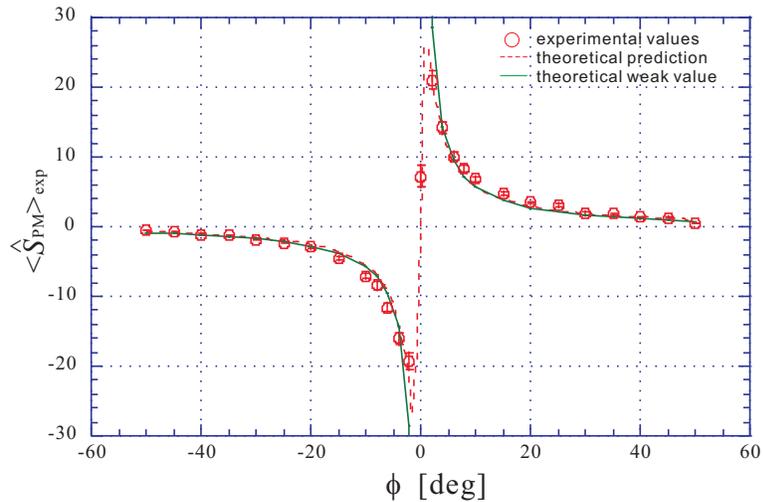}}
\caption{Experimental results of the weak measurement. The experimental weak values $\langle \hat{S}_{\mathrm{PM}} \rangle_{\mathrm{exp}}$ are shown 
as a function of initial polarization angle $\phi$. The open circles indicate the experimental data obtained from the conditional probabilities of the post-selected results, the broken line shows the expected effects of back-action given by Eq.(\ref{eqn:exp-wv}), and the solid line shows the theoretical weak value.}
\label{fig:result1}
\end{figure}

Fig. \ref{fig:result1} shows the experimental results for the weak values obtained with the measurement resolution obtained by setting the HWPs to $\theta = 0.5^\circ$. Theoretically, this corresponds to a measurement resolution of $\epsilon_{\mathrm{PM}}=0.035$ and a back-action related transition probability of $\eta_{\mathrm{HV}}=0.0003$. The experimental results are in good agreement with the theoretical weak values up to and including the measurement values obtained at $\phi = \pm 4^\circ$. The three measurement values obtained close to $\phi=0$ are consistent with the theoretical prediction for the back-action effects given by Eq.(\ref{eqn:exp-wv}) for $\eta_{\mathrm{HV}}=0.0003$. This correspondence suggests that almost all the flips in HV polarization are caused by the rotation of the HWPs in the measurement setup, with only negligible contributions from additional error sources.

The extremal weak values observed in the experiment were found at $\pm 20$. Since the input angles for these values are at $\phi=\pm 2^\circ$, this is lower than the maximal value of $\pm 28.6$ theoretically predicted for angles of about $\phi=\pm 1^\circ$. However, even the achievement of a 20 fold enhancement of the weak value requires a transition probability below $0.0006$. If the weak measurement was realized by a separation of the P and M polarizations followed by an interference between the paths to partially erase the measurement information, the visibility of the interference needed to obtain a 20 fold enhancement of the weak value would have to be as high as $\mathrm{V_{PM}}=1-2\eta_{\mathrm{HV}}=0.9988$. It is therefore essential that our setup only uses interferences between the H and V polarized paths, avoiding the errors that would be introduced by limited visibilities in path interferences between P and M polarization.

In our setup, the visibility of the path interference between the H and V polarized components was found to be $\mathrm{V_{HV}}=0.71$. The effects of this error reduce the measurement resolution by introducing transitions between the P and M polarizations. As a result, the measurement resolution is  $\epsilon_{\mathrm{PM}}=0.025$ instead of the ideal value of 0.035 predicted from $\theta$. However, this reduced resolution has no effects on the observation of weak values at low $\theta$, since weak values are always obtained from averages over a sufficiently high number of low resolution measurements. 
The experimental results thus confirm the main merit of our method for the realization of weak measurements. Oppositely, the method is not as suitable for strong measurements, where back-action is always maximal and an optimization of measurement resolution is desirable. Since we can vary the measurement strength of our setup continuously between weak and strong measurements, we can illustrate the performance of our setup in these very different operating regimes in terms of the experimental errors observed in measurement resolution and back-action as the measurement strength is varied by rotation the HWPs from $\theta=0^\circ$ to $\theta=22.5^\circ$. 

\section{Relation between measurement resolution and back-action}
\label{sec:urel}

In principle, the measurement resolution $\epsilon_{\mathrm{PM}}$ and the measurement back-action given by $\eta_{\mathrm{HV}}$ should be defined in terms of the experimental input-output relations of the measurement setup. For a specific input state with a PM polarization of $\langle \hat{S}_{\mathrm{PM}}\rangle=2 \mathrm{Re} [C_{\mathrm{H}}^* C_{\mathrm{V}}^*]$, the measurement resolution is given by the ratio between the output probability difference and the expectation value of the Stokes parameter in the input,
\begin{equation}
\epsilon_{\mathrm{PM}}=\frac{P(\mathrm{b1})-P(\mathrm{b2})}{2 \mathrm{Re} [C_{\mathrm{H}}^* C_{\mathrm{V}}^*]}, 
\end{equation}
where $P(\mathrm{b1})$ and $P(\mathrm{b2})$ are obtained from the total number of counts in $b1$ and $b2$. 
Likewise, the measurement back-action flips H and V polarization, reducing the input HV polarization of $\langle \hat{S}_{\mathrm{HV}}\rangle=|C_{\mathrm{H}}|^2-|C_{\mathrm{V}}|^2$ by a factor of $1-2 \eta_{\mathrm{HV}}$. If a measurement of HV polarization is performed in the output, the experimental measurement back-action is obtained from 
\begin{equation}
1 - 2 \eta_{\mathrm{HV}}=\frac{P(\mathrm{H})-P(\mathrm{V})}{|C_{\mathrm{H}}|^2-|C_{\mathrm{V}}|^2}, 
\end{equation}
where $P(\mathrm{\mathrm{H}})$ and $P(\mathrm{V})$ are the total H and V polarized counts summed over both $b1$ and $b2$. For consistency, it is convenient to define the measurement back-action as $2 \eta_{\mathrm{HV}}$, since a 
complete randomization of HV polarization ($P(H)=P(V)$) then corresponds to a back-action of $1$. 

In the absence of experimental errors, our measurement setup would have a measurement resolution of $ \epsilon_{\mathrm{PM}} = \sin 4 \theta $ and a back-action given by $1-2 \eta_{\mathrm{HV}} = \cos 4 \theta$, depending on the angles $\theta$ of the HWPs. This result achieves the uncertainty limit for resolution and measurement back-action in two level systems \cite{Englert96}, as given by the uncertainty relation  
\begin{equation}
\epsilon_{\mathrm{PM}}^2 + (1-2 \eta_{\mathrm{HV}})^2 \leq 1.
\label{eqn:uncertainty}
\end{equation}
In the actual experiment, linear decoherence effects reduce the values of $\epsilon_{\mathrm{PM}}$ and $1-2\eta_{\mathrm{HV}}$ from their ideal values to values below the uncertainty limit. If these reductions are expressed in terms of experimental visibilities, $ \epsilon_{\mathrm{PM}} = \mathrm{V_{HV}} \sin 4 \theta $ and $1-2 \eta_{\mathrm{HV}} = \mathrm{V_{PM}} \cos 4 \theta $, the actual relation between back-action and resolution can be described by 
\begin{equation}
\frac{\epsilon_{\mathrm{PM}}^2}{\mathrm{V_{HV}}^2} + \frac{(1-2\eta_{\mathrm{HV}})^2}{\mathrm{V_{PM}}^2} = 1.
\label{eqn:relation}
\end{equation}
If the values obtained for $\epsilon_{\mathrm{PM}}$ and $2 \eta_{\mathrm{HV}}$ are shown for different measurement strengths $\theta$, they should therefore lie on an ellipse around ($\epsilon_{\mathrm{PM}}=0, \, 2 \eta_{\mathrm{HV}}=1$), where $\mathrm{V_{HV}}$ determines the resolution in the strong measurement limit at $\theta=22.5^\circ$, and $\mathrm{V_{PM}}$ determines the back-action in the weak measurement limit at $\theta=0^\circ$. 

\begin{figure}[h]
\centerline{\includegraphics[width=100mm]{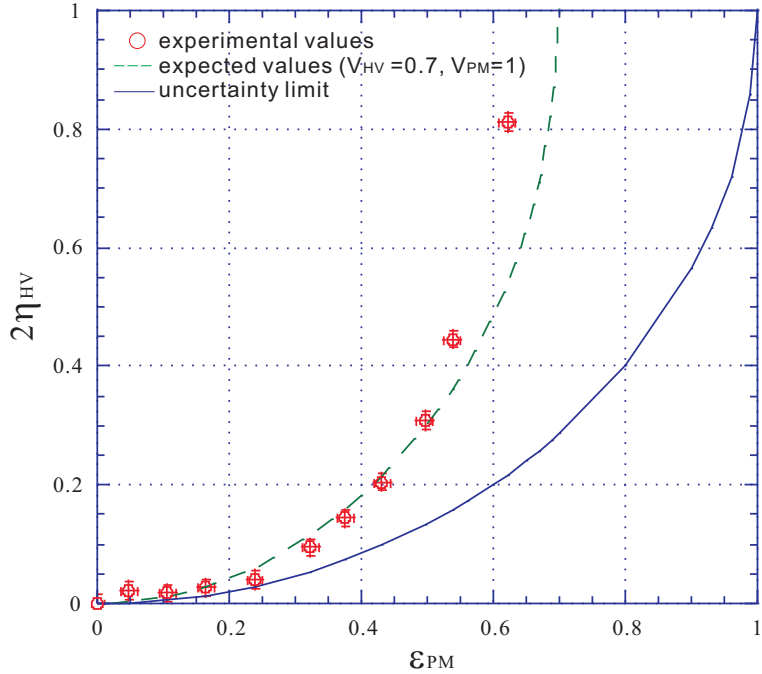}}
\caption{Relation between measurement back-action $\eta_{\mathrm{HV}}$ and measurement resolution $\epsilon_{\mathrm{PM}}$. 
Open circles indicate experimental values obtained for different measurement strengths $\theta$. The broken line shows the relation expected for a visibility of $0.7$ in our setup. The solid line indicates the uncertainty limit given by Eq.(\ref{eqn:uncertainty}).}
\label{fig:result2}
\end{figure}

Fig. \ref{fig:result2} shows the experimental results obtained with an input state at $\phi = 25^\circ$. The results reproduce the relation between resolution and back-action expected for $\mathrm{V_{HV}} = 0.7$ and $\mathrm{V_{PM}} = 1$ as shown by the broken line in the graph, except for some discrepancy in the values obtained in the strong measurement limit. Since the strong measurement limit is very sensitive to the visibilities of our interferometer, it is possible that these discrepancies may have been caused by instabilities in the interferometer. 

In general, the result is consistent with the values of $\mathrm{V_{HV}}=0.71$ and $ \mathrm{V_{PM}}>0.9988$ estimated from the weak measurement results. Since there is no experimentally resolvable limitation to the reduction of measurement back-action at low $\theta$,  the experimentally obtained relation between resolution and back-action confirms that our setup is particularly suited for weak measurements. Fig. \ref{fig:result2} thus illustrates the specific feature of our measurement setup in terms of the noise characteristics at different measurement strengths. 

\section{Conclusions}
\label{sec:conclusion}

We realized a weak measurement of diagonal (PM) photon polarization by path interference between the H and V polarized components. In this case, the visibility of the path interference depends on the amount of back-action induced by gradually rotating the orthogonal polarizations of the paths towards each other. It is therefore possible to control the amount of back-action precisely, while errors caused by the limited visibility of the path interference only affect the measurement resolution. This situation is ideal for the realization of weak measurements, since the achievement of extremely high weak values depends critically on the limitation of the total back-action to error rates below the post-selection probability. 

Our results show that we can achieve 20 fold enhancement of the weak values, even though the visibility of the path interference was only $0.71$. This robustness against experimental errors can be achieved because the measurement resolution is not relevant for the measurement of weak values. The requirements for operating in the weak measurement regime are therefore quite different from the requirements for operating in the strong measurement regime. We have characterized this difference in the experimental requirements by measuring the resolution and back-action of our setup at different measurement strengths. The present setup achieves the uncertainty limit in the weak measurement regime, but not in the strong measurement regime, where its measurement resolution is limited by the visibility of path interference. The characterization of errors for different measurement strengths thus confirms the specific usefulness of our approach for weak measurements. 

The setup presented here is easy to realize and allows the observation of extreme weak values even in the presence of significant experimental imperfections. It may therefore be useful in the characterization and control of quantum processes by weak measurements. In particular, it may greatly simplify the integration of weak measurements into optical quantum circuits, and the performance of multiple weak measurements in a cascaded system. We hope that these simplifications will help to establish weak measurements as part of the quantum information toolbox, leading to better insights into the fundamental properties of emerging quantum technologies.

\section*{Acknowledgements}
Part of this work has been supported by the Grant-in-Aid program of the Japanese Society for the Promotion of Science, JSPS.

\section*{Reference}


\end{document}